\documentclass[journal]{article}
\usepackage{amsfonts,amssymb,anysize}
\usepackage{times,mathptm}
\usepackage{amsmath,amsthm,graphicx}
\usepackage{hyperref}
\marginsize{0.9in}{0.9in}{0.6in}{0.8in}

\begin{document}

\title{Quantum Circuit Simplification and Level Compaction\thanks{
Copyright $\copyright$ IEEE.   Reprinted from IEEE Transactions on 
Computer-Aided Design of Integrated Circuits and Systems, 27(3):436--444, March 2008. \newline
This material is posted here with permission of the IEEE.  Internal or
personal use of this material is permitted.  However, permission to
reprint/republish this material for advertising or promotional purposes or
for creating new collective works for resale or redistribution must be
obtained from the IEEE by writing to
\href{mailto:pubs-permissions@ieee.org}{pubs-permissions@ieee.org}. \newline
By choosing to view this document, you agree to all provisions of the
copyright laws protecting it.}}

\author{Dmitri~Maslov\thanks{D. Maslov is with the Institute for Quantum Computing and Department 
Combinatorics and Optimization, University of Waterloo,
Waterloo, ON, N2L 3G1, Canada, email: \href{mailto:dmitri.maslov@gmail.com}{dmitri.maslov@gmail.com}.},
		Gerhard~W.~Dueck\thanks{G. W. Dueck is with the Faculty of Computer Science, University of New Brunswick,
Fredericton, NB, E3B 5A3, Canada.},
		D.~Michael~Miller\thanks{D. M. Miller is with the Department of Computer Science, University of Victoria,
Victoria, BC, V8W 3P6, Canada.},
		Camille~Negrevergne\thanks{C. Negrevergne is with the Institute for Quantum Computing, University of Waterloo,
Waterloo, ON, N2L 3G1, Canada.}
}


\maketitle

\begin{abstract}
Quantum circuits are time dependent diagrams describing the process of
quantum computation. Usually, a quantum algorithm must be mapped into
a quantum circuit. Optimal synthesis of quantum circuits is
intractable and heuristic methods must be employed. With the use of
heuristics, the optimality of circuits is no longer guaranteed. In
this paper, we consider a local optimization technique based on
templates to simplify and reduce the depth of non-optimal quantum
circuits.  We present and analyze templates in the general
case, and provide particular details for
the circuits composed of NOT, CNOT and
controlled-$sqrt$-of-NOT gates. We apply templates to optimize various
common circuits implementing multiple control Toffoli gates and quantum
Boolean arithmetic circuits. We also show how templates can be used to
compact the number of levels of a quantum circuit. The runtime of our
implementation is small while the reduction in number of quantum gates
and number of levels is significant.
\end{abstract}



\section{Introduction}        
Research in quantum circuit synthesis is motivated by the growing
interest in quantum computation \cite{bk:nc} and advances in
experimental implementations \cite{quant-ph/0004104, ar:s-k, qcr, www:quant}. In realistic devices, experimental
errors and decoherence introduce errors during computation. Therefore,
to obtain a robust implementation, it is imperative to reduce
the number of gates and the overall running time of an algorithm.
The latter can be done by parallelizing (compacting levels) the
circuit as much as possible.

Even for circuits involving only few variables, it is at present
intractable to find an optimal implementation. Thus a number of
heuristic synthesis methods have emerged. Application of these methods
usually results in a non-optimal circuit, that can be simplified with
local optimization techniques. Additionally, some quantum circuits for
important classes of functions, such as adders and modular
exponentiation, were created and compacted in an \emph{ad-hoc} manner
\cite{quant-ph/0406142, ar:mi}.

Local optimization has only recently been considered as a possible
tool for the gate count reduction in quantum \cite{quant-ph/0307111}
and reversible (quantum Boolean) circuits \cite{ws:iky}.  Some quantum
circuit identities that could be used for circuit simplification can
be found in \cite{bk:nc}. While these provide several rewriting rules
with no ready to use algorithm for their application,
there is clearly a benefit in a systematic approach through the use of
templates discussed in this paper.  A somewhat different approach for
local optimization of reversible NOT-CNOT-Toffoli circuits was applied
for the simplification of random reversible circuits in \cite{ws:sppmh}. 
That approach and our template method are difficult to compare as they 
have been applied to different types of circuits with different metrics 
for the circuit cost. 

So far, CAD tool designers spent little effort on minimizing the number of
logic levels in quantum circuits. However, this allows a shorter
running time as it results in a parallelization of the algorithm. 
More importantly, in the popular quantum error model where errors 
appear randomly with time, a parallel circuit helps to reduce the errors. 
For instance, it may be possible to use smaller number of error correction 
code concatenations (each of which is a very expensive operation, requiring
to at least triple the number of physical qubits \cite{bk:nc}) 
if the circuit is well parallelized. To
the best of our knowledge, all of the presently existing quantum
circuits were at best compacted in an \emph{ad-hoc} fashion.  In our
work we automate level compaction through the use of templates.

Methods based on templates have been considered for Toffoli
reversible network simplification \cite{ar:mdmCAD}.  In this
paper, we revisit the definition of templates and show how
they can be applied in the quantum case as a
systematic basis for the quantum circuit
simplification and level compaction.

The paper\footnote{A preliminary version of this work was presented
at DATE-2005 conference. This paper discusses circuit parallelization,
reports the improved results based on a new and significantly more efficient 
implementation, includes extensive testing results, as well as certain 
new discussions.} is organized as follows. We start with a
brief overview of the necessary background in Section
\ref{sec:back}. In Section \ref{sec:td} we define the templates and
discuss some of their properties. We
present a method to identify the templates and describe two algorithms,
one to reduce the cost and the other to reduce the
number of logic levels of quantum circuits in Section
\ref{sec:ta}. We next choose a specific quantum gate library, and
illustrate effectiveness of the above approach. Section \ref{sec:ncv}
presents a set of small quantum templates
for NOT, CNOT and controlled-$sqrt$-of-NOT gates and
illustrates the algorithms.  The benchmark results presented in
Section \ref{sec:res} are divided into two parts. We first optimize
quantum implementations of the multiple control Toffoli gates
(including multiple control Toffoli gates with negative controls) and
then consider optimization of some NOT-CNOT-Toffoli circuits available
through existing relevant literature.  Discussion of future work and
concluding remarks are found in Sections \ref{sec:fur} and
\ref{sec:conc}.

\section{Background}\label{sec:back}

We present a short review of the basic concepts of quantum
computation necessary for this paper. An in-depth coverage can be found in \cite{bk:nc}.

The state of a single qubit is a linear combination $\alpha |0\rangle
+ \beta |1\rangle$ (also written as a vector $(\alpha, \beta)$) in the
basis $\{|0\rangle,\;|1\rangle\}$, where $\alpha$ and $\beta$ are
complex numbers called the amplitudes, and
$|\alpha|^2+|\beta|^2=1$. Real numbers $|\alpha|^2$ and $|\beta|^2$
represent the probabilities $p$ and $q$ of reading the logic
states $|0\rangle$ and $|1\rangle$ upon measurement.  The state of a
quantum system with $n>1$ qubits is given by an element of the tensor
product of the single state spaces and can be represented as a
normalized vector of length $2^n$, called the state vector. Quantum
system evolution allows changes of the state vector through its
multiplication by $2^n \times 2^n$ unitary matrices called
gates.

The above models how a transformation can be performed, but does not
indicate how to identify the unitary operations 
that compose the transformation or how to implement
them. Efficiency of the physical implementation depends 
on the system's Hamiltonian, and the details of different systems 
(and associated gate costs) are not a focus of this paper. 
Typically, certain primitive gates are used as elementary
building blocks
\cite{bk:nc, ar:bbcd}. Among these are:
\begin{itemize}
\item NOT ($x \mapsto \bar{x}$) and CNOT ($(x,y)\mapsto (x, x \oplus
y)$) gates, where $x,y \in \{0,1\}$ and $\oplus$ is addition modulo 2;

\item Hadamard gate defined by
$H=\frac{1}{\sqrt{2}}\left(1\:\;\;\;1\atop 1\;-1\right)$;

\item controlled-V gate that depending on the value on its control
qubit changes the value on the target qubit using the transformation
given by the matrix ${\bf V}=\frac{i+1}{2}\left(1\;\;-i\atop
-i\;\;1\right)$;

\item controlled-$V^{+}$ that depending on the value of its control
qubit changes the value on the target qubit using the transformation
${\bf V^{\dagger}}={\bf V}^{-1}$;

\item rotation gates $R_a(\gamma)$, $\gamma \in [0,2\pi],\; a \in \{x,y,z\}$.
\end{itemize}

We shall write $G^{-1}$ to denote the gate implementing the
inverse function of the function realized by gate $G$. In context,
we will use $G$ to mean a gate or the transformation matrix for
that gate. The circuit diagrams are built in the popular notations, 
such as those used in \cite{bk:nc}. In short, horizontal ``wires'' 
represent a single qubit each; the time in the circuit diagrams is 
propagated from left to right; (positive) gate controls are depicted 
with $\bullet$; targets appear as $\oplus$ for NOT and CNOT gates, 
$\framebox{V}$ for controlled-$V$ gate, and $\framebox{V+}$ for
controlled-$V^+$ gate with vertical lines joining control(s) of 
a gate with its target.

The principle of the optimization method is to associate a
cost to each of these elementary gates and lower the overall circuit
cost by reducing the number of high cost gates. The cost definition
must reflect how difficult it is to implement the gate is and therefore will
depend on the details of the physical device considered to implement
the circuit. For example, for NMR techniques the cost of the gate must
take into account the number of rf-pulses as well as the duration of
the interaction periods necessary to implement the gate \cite{quant-ph/0004104}.
In a setting guided by the Ising type Hamiltonian in a weak coupling regime 
(such as liquid NMR \cite{quant-ph/0004104} and superconductors \cite{quant-ph/0603224})
a controlled-$V$ and its complex conjugate must be associated with approximately half 
the cost of a CNOT gate each. Thus, controlled-$V$ and controlled-$V^{+}$ are 
not at all complex gates. Two qubit gate implementation costs in any given Hamiltonian 
can be found using the technique discussed in \cite{ar:zvs}. 

The Toffoli gate \cite{ar:tof} and its generalization with more than
two controls serve as a good basis for synthesis purposes. Indeed,
every reversible (quantum Boolean) function can be realized as a cascade of multiple
control Toffoli gates \cite{ar:bbcd, ar:mdmCAD}.
The multiple control Toffoli gate flips the target bit if the
control bits are in a given Boolean state.  Unfortunately, multiple
control Toffoli gates (including the original Toffoli 
gate \cite{ar:tof}) are not simple transformations in quantum
technologies. They require a number of elementary quantum
operations and Toffoli gates with a large number of controls can be
quite costly \cite{ar:bbcd}.  However, they can be implemented using
circuits composed with 3-qubit Toffoli gates \cite{ar:bbcd}. Finally,
the 3-qubit Toffoli gate can be constructed from a set of gates that
includes the NOT, CNOT, controlled-$V$ and controlled-$V^+$. We
therefore consider all these gates in Section \ref{sec:ncv} when we search for
templates to simplify the best known quantum circuits implementing
large Toffoli gates and reversible functions. In addition, any 
unitary can be synthesized as a generic quantum circuit through 
exploring the properties of matrix decompositions 
\cite{ar:bbcd, quant-ph/0406176, quant-ph/0504100}. We {\em do 
not} consider those circuits here, but point out that our 
circuit simplification techniques are applicable in any of 
the above cases.

\section{Templates: definition}\label{sec:td}

To decrease the cost of a circuit, the basic idea is to replace a
sub-circuit with an equivalent one that has lower
cost.  We will call this procedure the application of a {\bf rewriting
rule}.  Some problems arise with this technique:

\begin{itemize}
\item{In general, even for simple circuits, if rotation gates
with any parameter $\gamma$ are allowed the number of possible
rewriting rules is infinite.}

\item{Equivalent circuits with same cost might require different sets
of rewriting rules to be simplified.}

\item A sub-circuit may be rewritten in another form having the same
cost, but this second form could allow extra-simplifications
on the circuit using other rewriting rules.
\end{itemize}

One of the problems arising from these considerations is to
minimize the number of rewriting rules by keeping only ``essential''
ones.  To address these issues we
introduce the notion of templates that will be applicable to all
quantum gate libraries and discuss the algorithms for quantum gate
reduction and level compaction.

\vspace{1mm}
\noindent {\bf Definition:} A {\bf size $m$ template} is a
sequence of $m$ gates that implements the identity operator and that
satisfies the following constraint: any template of size $m$ must be
independent of all templates of smaller or equal size, {\it i.e.}  for
a given template $T$ of size $m$ no application of any set of
templates of smaller or equal size can decrease the number of gates in
$T$ or make it equal to another template.

A {\bf template} can be seen as a generalization of the rewriting
 rules, since rewriting rules can be derived from it. For example, {\bf forward
 application} of the template $G_0\;G_1...\;G_{m-1} = I$
 allows us to find a rewriting rule of the form $G_iG_{(i+1) \bmod
 m}...\;G_{(i+p-1) \bmod m} \rightarrow G^{-1}_{(i-1) \bmod m}$
 $G^{-1}_{(i-2) \bmod m}...\;G^{-1}_{(i+p) \bmod m}$, where $0 \leq
 i,\;p \leq m-1$.
Similarly, {\bf backward application} of the template is a
rewriting rule of the form $G^{-1}_iG^{-1}_{(i-1) \bmod
m}...\;G^{-1}_{(i-k+1) \bmod m} \rightarrow G_{(i+1) \bmod m}$
$G_{(i+2) \bmod m}...\;G_{(i-k) \bmod m}$, where $0 \leq i,\;p \leq
m-1$.
\vspace{1mm}

Template application requires that the inverse of each gate be
available. Clearly, templates are a more compact way of representing non
redundant rewriting rules as it is capable of storing up to $2m^2$
rewriting rules.

See Appendix for a proof of the effect of the forward and backward 
applications of templates.

\section{Templates: application}\label{sec:ta}

In this section we present a method to find and classify the templates
and introduce two algorithms using them. One is
an algorithm for quantum cost reduction and the other for
quantum circuit level compaction, both based on the notion of the
templates.

\subsection{Template identification}

First we find all templates of the form $AA^{-1}$ (length 2),
which we call {\bf gate-inverse rules}. This is straightforward, 
since every self-inverse gate $A$ forms the template $AA$ and
every pair of gates $A$ and $B$, where $B=A^{-1}$ forms one template
of the form $AB$.
 
Subsequent templates are found by identifying increasingly longer
sequences of gates that realize the identity function and that
can not be reduced by other available templates.

Templates of the form $ABAB$ (length 4) with $A=A^{-1}$ and $B=B^{-1}$ 
applied for parameter $p=2$
result in construction of the rewriting rules $AB \rightarrow BA$ and
$BA \rightarrow AB$. That is, they define the conditions under which two
gates commute. We call such templates {\bf moving
rules} and apply them to move gates to form matches leading to reduction
via other templates.

For applications, we suggest seeking a complete classification of the
templates of small size and then supplementing those by a set of
templates that appear to be useful when a specific synthesis procedure
is applied. For example, if a synthesis procedure (or the circuit
types one considers) tends to use a specific type of sub-circuit of
cost $\mu$ which is neither optimal (assume an optimal cost of $\nu$)
nor can be simplified by a small size complete set of templates, a
template with total cost $\mu+\nu$ can be created 
(followed by a generalization process when and if needed). In this
paper, we do not construct any of these supplementary type templates,
since we apply templates to the
circuits from different authors obtained from different synthesis
procedures.

\subsection{Cost reduction}

In this section, we present an algorithm to reduce \emph{generic}
quantum circuit cost using the templates.  To apply the algorithms to
a \emph{specific} physical implementation, we only need to
choose a relevant cost definition.\\

\noindent {\bf Input:} A quantum circuit specification, {\em i.e.} a
sequence of gates $C_1 C_2 ... C_n$.\\

\noindent {\bf Output:} A quantum circuit computing the same
function as the input circuit, but having a possibly lesser
cost.\\

\noindent {\bf Algorithm:} 

\begin{enumerate}

  \item Let $C_k$ be the {\em start} gate in the circuit for a
  potential template match.  Initially $k=2$.

  \item {We attempt to match
   the templates in order of size (excluding the
  moving rules).  The attempt to match to a size $m$ template
  $G_0G_1... G_{m-1}$ proceeds as follows:

	\begin{enumerate} \label{first1}

	  \item{ {\em Forward matching:} Apply the moving
	    rules to arrange the gates preceding $C_k$ to be able to
	    match them with the given size $m$ template. At this step we determine
	    pair $(j,p)$ such that $C_{k-i}=G_{(j+i)\bmod m},\;0\leq
	    i<p$.  When such a $j$ and $p$ are found, gates
	    $C_{k-p-1},C_{k-p-2},...,C_k$ can be replaced by the
	    sequence $G^{-1}_{(j+p+i)\bmod m},\;0\leq i<m-p$.
	    Substitution is done if it is {\em beneficial} from the point of 
	    view of the overall circuit cost reduction.}
	    
	  \item{{\em Backward matching:} To backward match a size
	    $m$ template, the same procedure applies with the following
	    matching condition: $C_{k-i}=G^{-1}_{(j-i)\bmod m},\;0\leq
	    i<p$. Then, gates $C_{k-p-1},C_{k-p-2},...,C_k$ can be
	    replaced by the sequence $G_{(j+p-i)\bmod m},\;0\leq
	    i<m-p$. The decision to replace or not is based on a chosen
	    circuit cost metric.}
	 \end{enumerate}
}

  \item We propagate this procedure through the circuit:
    \begin{itemize}

    \item If a template substitution was made, then $k$ is set to
    index of the leftmost gate substituted and we repeat step \ref{first1}.

    \item Otherwise, if we can, increment $k$ by 1 and
	repeat step \ref{first1}.  If we cannot because $C_k$ is
	already the rightmost gate in the circuit, the algorithm 		
	terminates.

    \end{itemize}

\end{enumerate}

The gate replacement at step \ref{first1} is performed when it is {\em
beneficial} to do such replacement {\it i.e.} when the total circuit
cost is reduced. This imposes extra constraints on the parameter $p$
depending on the exact cost definition. For instance, with simple gate
count cost metric, $p$ must be greater than $m/2$. If many pairs $(j,p)$
are found, the one associated to the biggest cost reduction is chosen for 
the gate substitution. However, even if the total cost after template
application stays the same (for simple gate count cost metric this means
applying an even size template by replacing its half with another half,
{\it i.e.} for even $m$ and $p=m/2$) the substitution can be beneficial
as the new circuit arrangement may allow other cost reducing template
applications. We take this into account by allowing such ``cost
retaining'' template applications as long as $k < Flag$ ($k$ is value of
the subscript of $C_k$), with the $Flag$ initially set to $0$. After
each cost retaining template application the $Flag$ is set to the
current $k$ value, and after each cost reducing template application the
$Flag$ is set back to $0$. This guarantees that the cost reduction
algorithm will not run into an infinite loop while allowing cost
retaining template application.

In Section \ref{sec:ncv} we illustrate how the templates are applied
to reduce the gate count.

\subsection{Level compaction}

We next suggest a greedy algorithm
for quantum circuit level compaction employing templates. A level is
defined as a sub-sequence of commuting gates that can be applied in
parallel. Level compaction
helps to increase the parallelization of the circuit implementation and
therefore not only optimizes the runtime of the circuit, but also helps to 
decrease the decoherence effects by shortening the overall execution 
time\footnote{For instance, a liquid NMR circuit with a high degree of
parallelization of single qubit rotations and ZZ-gates will be
singnificantly shorter than its unparallelized version.  Indeed, single qubit
rotations on homonuclear spins are usually implemented by selective
soft pulses sent sequentially to act on each spin. Nevertheless, if
we want to act on all homonuclear spins in parallel, it is possible to 
use a single broad-band short pulse \cite{quant-ph/0004104}. As
for heteronuclear spins, modern spectrometers have several channels that 
can be used simultaneously. Therefore, one can rotate heteronuclear 
spins in parallel by pulsing on them in parallel.

More importantly, in a typical NMR system, the main time consuming
gates are the interaction gates (ZZ gates). Because all the couplings
are always on in a molecule, ZZ-gates naturally occur in parallel in
the circuit. To apply a ZZ gate to a given pair of qubits, one needs to
use refocussing techniques \cite{quant-ph/0004104} involving pulses
and delays to cancel all the ZZ-interactions but the desired one.
Therefore, in most of the cases, regrouping the ZZ-gates will allow to
optimize the refocussing scheme and reduce the overall number of
required delays. In particular, refocussing scheme exists for any subset 
of non-intersecting gates defined as a single logic level in this paper \cite{ar:jk}.}.
For {\em simplicity} of the algorithm description, we assume 
that all gates have the same duration, therefore, the execution time
of a level is equal to a single gate duration. We also assume that
neighboring gates operating on disjoint qubit
subsets can always be applied in parallel, which is a common 
assumption for quantum technologies.\\

\noindent {\bf Input:} A quantum circuit specification, {\em i.e.}
a cascade of gates $C_1C_2...C_n$.\\

\noindent {\bf Output:} A re-organized circuit with possibly fewer levels
computing the same function as the input circuit.\\

 \noindent {\bf Algorithm:} The principle is to assign a
 specific level to each gate.
 \begin{enumerate}
\item Initially, all gates in the circuit have undefined level, $i=1$
and we define $Qlevel_i$ as an empty set.
 
\item \label{first2} Consider $C_j$ the leftmost gate not yet assigned a
 level.  Assign it level $i$.
 
\item \label{last2} Until each gate $C_k$ right of $C_j$ is considered:
\begin{enumerate}
    \item If gate $C_k$ does not share common 
    qubits with any of the gates in level $Qlevel_i$ and gate $C_k$ can be moved left (using the moving rules)
    until it is adjacent to the leftmost gate with level $i$
    and then assign gate $C_k$ level $i$.
    \item If it is not possible to move gate $C_k$ as just
    described, apply templates using the algorithm described
    above using $C_k$ as the start gate and considering only those gates
    whose level has not been assigned yet.  Only templates with an
    even number $m$ of gates are applied and only substitutions of
    $m/2$ for $m/2$ gates are made.  Such substitution may
    allow a gate to subsequently (possibly with
    movement) to be assigned to level $i$.
\end{enumerate}
\item If there still remain gates not assigned a level, add 1 to $i$
(the number of levels), consider new empty $Qlevel_i$ and repeat steps \ref{first2} to \ref{last2}.
 \end{enumerate}

At this stage of development, the level compaction algorithm
is greedy. We expect that it can likely be improved. However, our
tests have shown that its current performance already 
improves relevant quantum circuits.

\section{Quantum NCV templates}\label{sec:ncv} 

\begin{figure}
\centering
\includegraphics[height=36mm]{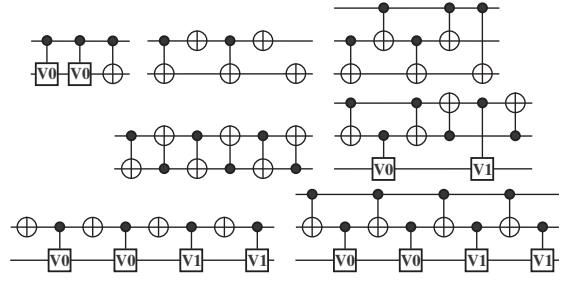}
\caption{Quantum templates other than the gate-inverse and moving
rules. Each of these circuits implements the Identity.}
\label{quant_temp}
\end{figure}

We now present a set of quantum templates based on the NOT, 
CNOT and controlled-$sqrt$-of-NOT (NCV) gates. It contains:

\begin{itemize}

\item{ The {\em gate-inverse rules:} NOT and CNOT are self-inverses
and controlled-$V$ and controlled-$V^+$ are the inverses of each
other.}

\item{ The {\em moving rule} (replace $AB$ with $BA$): assuming gate $A$
has control set $C_A$ ($C_A$ is an empty set in the case of an
uncontrolled gate) and target $T_A$ and gate $B$ has control set $C_B$
and target $T_B$, these two gates form a moving rule if, and only if,
$T_A \not\subseteq C_B$ and $T_B \not\subseteq C_A$.}

\item{{ \em Larger templates:} all other templates that we have
identified are shown in Figure \ref{quant_temp}, where $V$
(alternatively $V^+$) is substituted for all occurrences of $V_0$ and
$V^+$ (alternatively $V$) is substituted for all occurrences of $V_1$,
{\it i.e.} the substitution is consistent and distinct for
$V_0$ and $V_1$.  The templates reported here were found by
inspection.  We are currently developing a program to
find larger templates systematically and to verify
completeness of the current set.}

\end{itemize}

\begin{figure}[t]
\centering
\includegraphics[height=44mm]{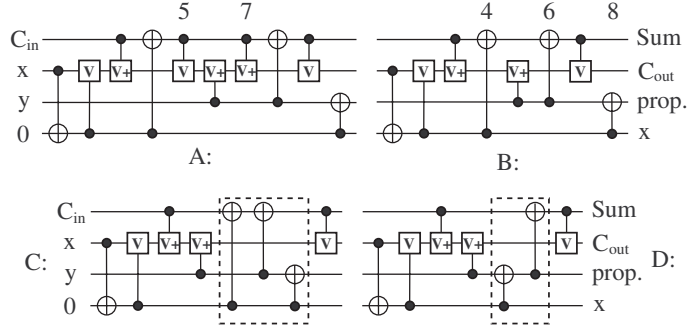}
\caption{Simplification of a 10-gate quantum network for the 3-qubit full adder.}
\label{quantadder}
\end{figure}

To illustrate how templates are applied, consider the quantum circuit
for the 3-input full adder with 10 gates from \cite{co:hsyyp}. The circuit
is built on four qubits as the 3-input adder must be extended to a 4-variable
reversible function. Note that the original circuit presented in
\cite{co:hsyyp} gives 1111 as output for input pattern 0100 instead of
the expected 1011. The circuit shown in Figure \ref{quantadder}A
corrects this.


In the circuit in Figure \ref{quantadder}A, gates 5 and 7
(counting from the left) can be moved together and form a
gate-inverse pair. We move them together and delete them by
applying the gate-inverse rule. 
This results in the circuit illustrated in Figure \ref{quantadder}B.

Next, we notice that gates 4, 6 and 8 in this circuit can also be
brought together (gates 4 and 8 should be moved towards gate 6.
Figure \ref{quantadder}C shows the three gates brought together, and
Figure \ref{quantadder}D illustrates the resulting circuit after the
size 5 template is applied.

The circuit that we found using templates simplification (Figure
\ref{quantadder}D) is equivalent to the optimal (for a given
input-to-output assignment) reported in \cite{co:hsyyp}. It took our
program $<0.001$ seconds (elapsed time on a 1.8 GHz Athlon XP2400+
machine with 512 MB RAM running Windows) to simplify the circuit in
Figure \ref{quantadder}A into the circuit in Figure \ref{quantadder}D.
The time reported in \cite{co:hsyyp} to synthesize such a circuit is 7
hours. This example clearly shows that templates are useful and
effective.

\begin{figure*}[t]
\centering \includegraphics[height=25mm]{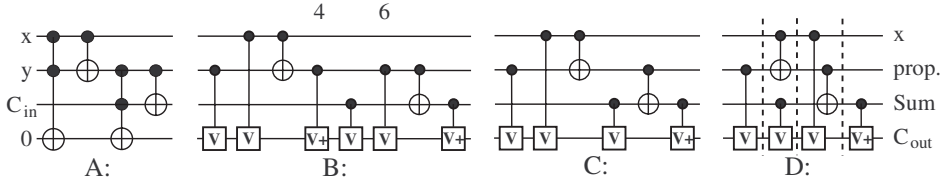}
\caption{Simplification of an 8-gate quantum circuit for the 3-qubit full adder.}
\label{adder}
\end{figure*}

A likely optimal quantum circuit for the 3-input full adder can be
constructed from its well-known reversible implementation illustrated
in Figure \ref{adder}A. We first substitute quantum circuits for the
Peres gates \cite{ar:peres} each of which is a Toffoli-CNOT pair (see Figure
\ref{adder}B). We then apply the templates. In this case, gates 4 and
6 can be moved together and match the gate-inverse rule. So, they are
both deleted leading to the circuit in Figure \ref{adder}C. Finally,
we apply the level compactor and the circuit is transformed into the
one illustrated in Figure \ref{adder}D (different logic levels are
separated by dashed vertical lines). The number of levels in the
compacted circuit is 4, and this is minimum due to having 4 gates with
targets on qubit $0-C_{out}$.

\subsection{Other templates}

It is possible to construct the templates in other gate libraries and then use the 
discussed cost reduction and level compaction algorithms verbatim. Constructing the templates 
for the finite (those seem to be more physical) gate libraries may be reduced 
to finding the rewriting rules by hand and generalizing them into the templates,
or running a computer search. A parameterization/classification of the templates 
in this case may be helpful. However, in the libraries with an infinite
number of gates a classification is necessary. We suggest that each template (template 
class) is written in the circuit form and followed by an algebraic expression 
conditional upon which the template applies. For example, in the library with 
single qubit rotations and CNOT gates the following template might be constructed 
$R_a(\alpha)R_a(\beta)R_a(\gamma)$, where $\alpha+\beta+\gamma=0$. Application of such 
template can be thought of as finding two single qubit rotations about the same axis (not 
necessarily the conventional $X$, $Y$ or $Z$, but a possible combination of them), 
that can be commuted until they are neighbors, and then they get replaced by a
single cumulative rotation. Another example of a template for this gate library could 
be $R_X(\alpha)R_Z(-\pi/2)R_Y(\alpha)R_Z(\pi/2)$, which could be used to 
replace some three gates with one, or, for instance, eliminate all $R_X$ 
gates from a given circuit. In the gate library with controlled gates, the following
template is possible $CU(b,c)CNOT(a,b)CU^{\dagger}(b,c)CNOT(a,b)CU(a,c)$, conditional
upon gate $CU$ being a self inverse. This template is a generalization of the 
one used in this paper (third template in Figure \ref{quant_temp}), but it 
captures an infinite number of the rewriting rules. Other templates are 
possible and depend on the gate base considered. The discussed examples 
are not intended to be treated as complete review of the possible templates,
rather an illustration what kind of templates may be constructed.

\section{Numerical results}\label{sec:res}

\begin{figure}[h]
\centering
\includegraphics[height=18mm]{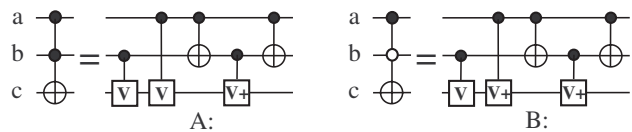}
\caption{Optimal NCV circuits for A: 3-qubit Toffoli gate \cite{bk:nc} 
and B: 3-qubit Toffoli gate with a single negative control ($b$) 
\cite{quant-ph/0511008}.}
\label{toffolis}
\end{figure}

Reversible logic and quantum arithmetic circuits are often specified with NOT, CNOT and Toffoli gates 
\cite{ar:bbcd, quant-ph/0406142, wp, ar:mi, bk:nc}, rather than with gates from 
the NCV set. Circuits with multiple control Toffoli gates have 
been studied extensively and synthesis procedures exist.
To process these circuits we need to transform every Toffoli gate
into a circuit with NOT, CNOT, controlled-$V$ and controlled-$V^+$
gates. We use the circuit in Figure \ref{toffolis}A for this purpose.
Due to the symmetry properties of the NCV and Toffoli gates
(interchangeability of the controlled-$V$ and controlled-$V^+$ gates
in quantum NCV circuits for reversible functions
\cite{quant-ph/0511008}, symmetry of Toffoli gate controls, and
self-inverse property of the Toffoli gate),
there exist 8 distinct but equivalent NCV circuits for a Toffoli gate.
In our procedure we use only two of them: the circuit in Figure
\ref{toffolis}A and its inverse, and keep the one resulting in
a better circuit simplification. Empirical test have shown that the use 
of other six transformations will not yield any new improvements.

\subsection{Multiple control Toffoli gate simulations}

Multiple control Toffoli gates and their variants with negated controls 
are a popular basis for the
synthesis of reversible circuits and are often used to construct
quantum circuits.  For instance, multiple control Toffoli gates are used 
in quantum error correcting circuits right after the syndrome was found to correct 
errors \cite{bk:nc}. Even more importantly, multiple control Toffoli gates 
are at the heart of amplitude amplification technique \cite{quant-ph/0005055} that 
is often considered as a separate class of quantum algorithms, of which there 
are only a few. Thus, multiple control Toffoli gates are indispensable for quantum 
computations, and it is very important to have efficient quantum
circuits for them.  Implementations
of multiple control Toffoli gates were studied in 
\cite{quant-ph/0512016, ar:bbcd}. In the following, we simplify and compact levels
in the multiple Toffoli gate circuits described in \cite{quant-ph/0512016, ar:bbcd} using our
templates based algorithms. We compare our results to
those presented initially. Table \ref{tab:tof} summarizes the results.

\begin{table}
\begin{center}
\caption{Simplification of the multiple control Toffoli gate 
implementations by Barenco {\em et al.} \cite{ar:bbcd} and 
Asano {\em et al.} \cite{quant-ph/0512016}. The results are 
grouped in two tables according to source of the initial 
circuit. Columns {\em Size} and {\em Ancilla} show
the size ($n$-qubit gate) of the multiple control Toffoli
gate, and the number of ancilla qubits associated with the
implementation of this gate. Columns {\em [citation] GC} and
{\em [citation] D} present the gate count (GC) in the best reported
quantum NCV circuit taken from the appropriate source indicated 
in {\em [citation]}, and the corresponding circuit depth (D). 
We show the gate counts and circuit depth for our optimized
implementations in columns {\em Opt-d GC} and {\em Opt-d D}. 
Whenever columns {\em [citation] D} and {\em Opt-d D} are not present this means 
that the depth equals to the number of gates both in the circuit 
before optimization and in the circuit after optimization.}
\vspace{2mm}
\begin{tabular}{|c|c|c||c|}\hline
Size          & Ancilla  & \cite{ar:bbcd} GC  & Opt-d GC \\ \hline
4             & 1        &      20  & 14         \\
5             & 2        &      40  &  26          \\
6             & 3        &      60  &  38          \\
7             & 4        &      80  &  50          \\
8             & 5        &     100  &  62           \\
9             & 6        &     120  &  74          \\
10            & 7        &     140  &  86           \\
11            & 8        &     160  &  98           \\
12            & 9        &     180  &  110          \\ \hline
$n>3$         & $n-3$    & $20n-60$ &  $12n-34$  \\ \hline
\end{tabular}

\vspace{2mm}

\begin{tabular}{|c|c|c|c||c|c|}\hline
Size & \!Ancilla\!  &  \cite{quant-ph/0512016} GC  & \cite{quant-ph/0512016} D & Opt-d GC & Opt-d D \\ \hline
6    & 3        &    80  & 40     & 60           & 29    \\
10   & 7        &   320  & 80     & 248          & 59    \\ \hline
$\!\!n$=$2^{m+1}+2\!$ & $n-3$ & $\!5(n-2)^2\!$ & $\!10n-20\!$ & $\!3.75(n-2)^2\!$ & $\!7.5n-10\!\!$ \\	\hline
\end{tabular}

\label{tab:tof}
\end{center}
\end{table}

The results in Table \ref{tab:tof} show that the set of multiple
control Toffoli gates of size $n$ realizations with gate count
of $20n-60$ (Lemma 7.2 in \cite{ar:bbcd}) are always simplified to the
circuits with $12n-34$ gates.  Based on the regularity and predictability of this
simplification, we conjecture this will always be the case. Further,
our experiment showed that asymptotically we get a $40\%$ reduction in
the number of gates and in the number of logic levels required in
simulation of the multiple control Toffoli gates. Similarly, 
circuits for multiple control Toffoli gates with $5(n-2)^2$ gates 
and depth $10n-20$ seem to always simplify to the circuits with 
$3.75(n-2)^2$ gates and having depth $7.5n-10$.

Multiple control Toffoli gates can be
implemented with a single auxiliary qubit as
discussed in Corollary 7.4 \cite{ar:bbcd}. Using our tool we achieved {\em upper bound}
$24n-88$ (for $n>5$) for the number of gates and the
number of levels required in multiple control Toffoli gate simulations
with a single auxiliary qubit using just the decomposition from Lemma 7.2 of \cite{ar:bbcd}. 
We stress that the above formulas are
{\em upper bounds} since we did not yet apply our techniques to
simplify such circuits. There must be a clever approach in which both 
types of $n-3$ auxiliary qubit decompositions are used in the construction due to 
Corollary 7.4 \cite{ar:bbcd}, and depending on whether
the final gate count or depth needs to be optimized, the choice
for a particular multiple Toffoli gate substitution may vary.

Multiple control Toffoli gates with negations
may also be useful in some applications.  A
canonical implementation of such gates (\cite{bk:nc} Figures 4.11 and
4.12) assumes a logic layer of NOT gates preparing the literals in the
right polarity followed by a multiple control Toffoli gate with all
positive controls and a level of NOT gates returning the values of
literals to the positive polarity. This makes multiple control Toffoli
gates with negative controls marginally more expensive than the multiple
control Toffoli gates with only positive controls. In the following, we
show that a multiple control Toffoli gate with some but not all negative
controls can be implemented with the same cost as a multiple
control Toffoli gate of the same size with only positive controls.

\begin{figure}[h]
\centering
\includegraphics[height=42mm]{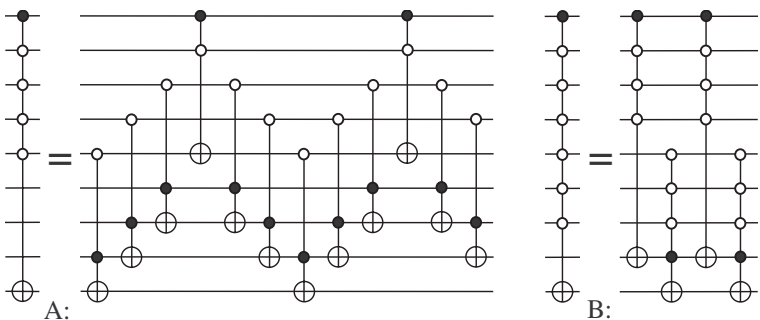}
\caption{Barenco {\em et al.} \cite{ar:bbcd} inspired 
simulations of multiple control Toffoli gate with 
some but not all negations illustrated for the maximal number of 
negative controls possible A: using $n-3$ auxiliary qubits  
and B: using a single auxiliary qubit.}
\label{lt}
\end{figure}

Given that the 3-qubit Toffoli gate with a single negated
control can be implemented with the same
(minimal) number of gates as a 3-qubit Toffoli gate with positive
controls \cite{quant-ph/0511008} (see Figure \ref{toffolis}B), such
gate can be used in the circuit proposed by Barenco {\em et al.}
\cite{ar:bbcd} to implement multiple control
Toffoli gates with some but not all negations with no cost
overhead. Such simulation is illustrated in Figure \ref{lt}A. 

Furthermore, such multiple control Toffoli gate with some but
not all negative controls implementations (\cite{ar:bbcd}, Lemma 7.2) 
rely on a similar strategy to simplify and
compact levels as the one used for multiple control Toffoli
gate with all positive controls. Therefore, each multiple control
Toffoli gate with some but not all negative controls
can be implemented with $(n-3)$ auxiliary
qubits, $12n-34$ CNOT, controlled-$V$ and controlled-$V^+$ gates, and
$12n-34$ logic levels $(n>3)$.  Using the simulation illustrated in
Figure \ref{lt}B one can construct multiple control Toffoli gate with
some but not all negations and requiring a single auxiliary qubit with
no more than $24n-88$ gates and the same number of logic levels (for
$n>5$).

Implementation of multiple control Toffoli gate with all negations
requiring $(n-3)$ auxiliary qubits will require 2 extra NOT gates,
however, the number of levels will not increase. Similarly, a multiple
control Toffoli gate with all negations simulation with a single
auxiliary qubit will require 4 extra NOT gates with no increase in the
number of logic levels ({\em upper bound}).

A similar argument holds in the case of the decomposition form 
\cite{quant-ph/0512016}, but we do not discuss this here. Rather, we move on to
considering other types of circuits. 

\subsection{Benchmark circuits}

Here we present the results of the application of the templates
to a number of quantum circuits implementing various reversible Boolean
and quantum arithmetic functions that can be found in the literature. Many
reversible/quantum circuits have constant input values and garbage
outputs. This typically occurs when a non-reversible function is
mapped to a reversible one prior to synthesis as a reversible circuit.
In such cases, extra simplifications at the extremities of the circuit can be
performed:
\begin{itemize}
\item{If a gate whose control is an input constant can be moved to the
  beginning of the circuit then depending on the constant input
  controlling the gate being 0 or 1 the gate can be either deleted or
  uncontrolled (assuming an uncontrolled gate has a lesser cost).}
  
\item{If a gate with the target on a garbage output can be moved to
  the end, we can delete it as we are not interested with the value
  of the garbage output.}
\end{itemize}

We took the circuits from \cite{wp} composed with NOT, CNOT, and 
Toffoli gates and compared their quantum realization costs 
(defined as NCV gate count) before and after applying the templates.
We also compacted levels in the simplified circuits and report the 
number of levels we get. Since \cite{wp} do not compact levels in 
their circuits, we have no comparisons for the number of levels. Table \ref{tab:bench} summaries the results.

\begin{table}
\begin{center}
\caption{Simplification of the benchmark circuits from \cite{wp}.
Circuit name appears in column {\em Name} and is taken directly from
\cite{wp}, {\em Size} indicates the number of qubits in the circuit.
{\em NCV GC} lists the quantum NCV gate count when the Toffoli gates
in the corresponding circuit are substituted with their quantum
implementations. {\em Optimized NCV GC} and {\em Levels} show
the quantum gate count and the number of logic levels after
reversible gates are substituted with their quantum
circuits and the resulting circuit is run through the template
simplification and then level compaction processes.  We do not report
the runtimes in this table because all circuits were computed almost
instantaneously.}
\vspace{2mm}
\begin{tabular}{|c|c|c||c|c|}\hline
Name & Size & NCV GC & Optimized NCV GC & Levels \\ \hline
$2of5d2$      & 7       & 40      & 29      & 25 \\
$rd32$        & 4       & 12      & 6       & 4 \\
$3\_17tc$     & 3       & 14      & 10      & 10 \\
$4\_49-12-32$ & 4       & 32      & 27      & 21 \\
$6symd2$      & 10      & 72      & 53      & 27 \\
$9symd2$      & 12      & 108     & 82      & 50 \\
$mod5d1$      & 5       & 24      & 14      & 9 \\
$mod5d2$      & 5       & 25      & 11      & 8 \\
$mod5mils$    & 5       & 13      & 9       & 5 \\
$ham3tc$      & 3       & 9       & 7       & 7 \\
$ham7-25-49$  & 7       & 49      & 40      & 28 \\
$hwb4-11-23$  & 4       & 23      & 21      & 16 \\
$rd53d2$      & 8       & 44      & 31      & 19 \\
$rd73d2$      & 10      & 76      & 55      & 34 \\
$rd84d1$      & 15      & 112     & 86      & 41 \\ \hline
\end{tabular}
\label{tab:bench}
\end{center}
\end{table}

\begin{figure}[t]
\centering
\includegraphics[height=52mm]{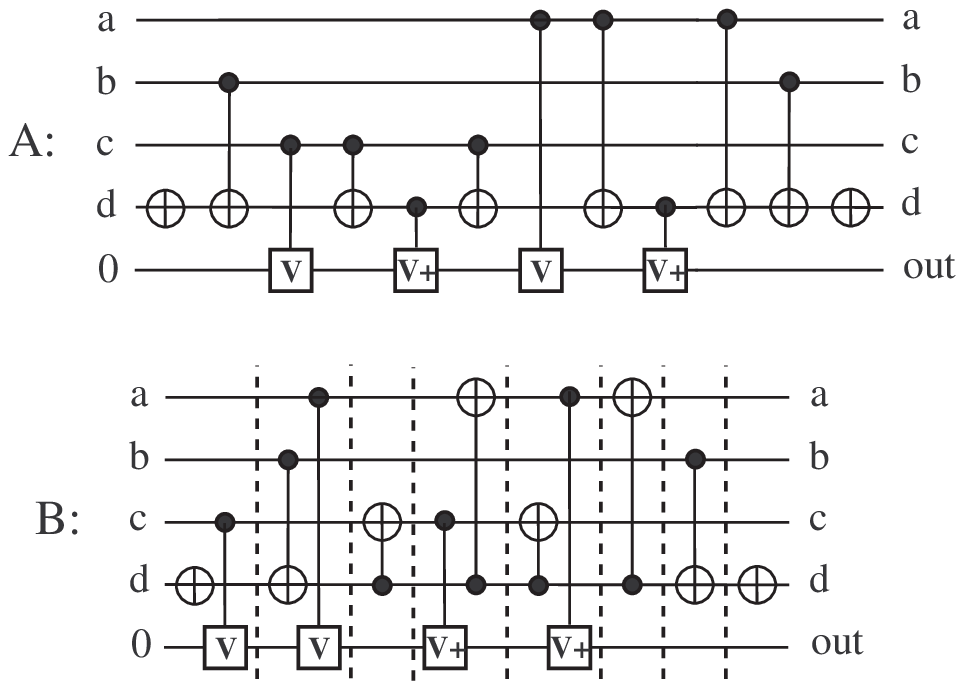}
\caption{Circuit for the oracle $mod5$.}
\label{grover}
\end{figure}

Let us describe the simplification procedure for one of these benchmark
circuits: the 5-qubit oracle function $mod5$. It leaves the first four
inputs unchanged and inverts the last one if, and only if, the first
four represent an integer divisible by 5. We first found a Toffoli gate
realization (circuit $mod5mils$ in Table \ref{tab:bench}). We then
applied the template based optimization techniques described above. The
resulting circuit is illustrated in Figure \ref{grover}A. If the inputs
are not required to be passed through unchanged, the last three gates
may be dropped. We next applied the level compaction algorithm. The
compacted version of the circuit in Figure \ref{grover}A is illustrated
in Figure \ref{grover}B. Note how the level compactor changes the form
of the circuit to allow fewer levels.  This happens when
even size templates are applied to change the form of the
circuit to facilitate further level compaction. 
Unless this is done, circuit in Figure \ref{grover}A cannot be compacted
to have less than 10 logic levels. This is because qubit $d$ is used 10
times as a control/target. If the inputs need not be recovered, the
depth of such computation is only 5 logic levels, and the number of
gates required is 9.

Finally, we applied the simplification procedure to some levelled
quantum circuits for adder, comparator and modular exponentiation type
function (the latter is an important part of the Shor's factoring 
algorithm) reported in \cite{quant-ph/0406142, ar:mi}.
We took their circuits, substituted
quantum implementations of the Toffoli gates where needed,
simplified them and compacted levels (treating each
circuit as non-levelled). In the circuit with Fredkin gates
(\cite{ar:mi}, Fig. 4) we used CNOT-Toffoli-CNOT decomposition of the
Fredkin gate, and in the circuit with single negative control Toffoli
gates (\cite{ar:mi}, Fig. 5), we used circuit from Figure
\ref{toffolis}B. The results are reported for 3 circuits that
can be found in \cite{quant-ph/0406142} and 3 circuits from
\cite{ar:mi} (Table \ref{tab:bench2}). 
 
\begin{table*}
\begin{center}
\caption{Simplification of the benchmarks from \cite{quant-ph/0406142, ar:mi}. 
{\em Name} shows where the
initial circuit can be found, {\em Size} lists the number of qubits
used, {\em NCV GC} lists the number of NCV gates required and {\em
Levels} shows the number of levels (each level with a Toffoli gate
considered to have width 5). Our results for the number of gates and
the number of levels are listed in columns {\em Optimized NCV GC} and
{\em Optimized levels}. The final column presents the total runtime
(elapsed time) required by our software to complete the circuit
simplification and compact the levels when run on an Athlon XP2400+
with 512M RAM machine under Windows.}
\vspace{2mm}
\begin{tabular}{|c|c|c|c||c|c|c|}\hline
Name & Size & NCV GC & Levels & Optimized & Optimized & Runtime \\ 
     &      &        &        & NCV GC    & levels    &  \\ \hline
\cite{quant-ph/0406142}, Fig. 5 & 35 & 368 & 86  & 303 & 53 & 1.883 sec\\ 
\cite{quant-ph/0406142}, Fig. 6 & 24 & 172 & 49  & 110 & 27 & 0.341 sec\\
\cite{quant-ph/0406142}, Fig. 7 & 26 & 337 & 101 & 287 & 61 & 1.903 sec\\
\cite{ar:mi}, Fig. 2 & 10 & 60  & 47 & 34  & 20 & 0.07 sec\\
\cite{ar:mi}, Fig. 4 & 15 & 70  & 44 & 58  & 23 & 0.210 sec\\
\cite{ar:mi}, Fig. 5 & 30 & 168 & 37 & 112 & 21 & 0.301 sec\\ \hline
\end{tabular}
\label{tab:bench2}
\end{center}
\end{table*}

\section{Future work}\label{sec:fur}

There are several possibilities to improve our simplification approach.
We are interested to develop a smart automated procedure for
substituting quantum circuits for multiple control Toffoli gates. The
search for the new templates can be accomplished finding all identities
of the given size and applying templates to simplify them. All
identities that do not simplify are the new templates. Such search
method is also suitable for proving the completeness of the set of the
templates found.

As far as level compaction goes, we presented a very simple and greedy
algorithm. We expect that our results for the number of levels can be
improved through use of a smarter level compaction algorithm. However,
we believe that the templates could still serve as an efficient core for
such improved level compactor.

Finally, we are interested in extending the experimental results of the
templates application to other sets of quantum gates including rotation
gates and elementary pulses (NMR quantum technology; this will be a
technology-specific optimization), and to 
account for different architectures (which should be straightforward since 
each undesirable gate can be punished with a high cost). Since the templates definition is
based on the properties of matrix multiplication only, they can be
applied in {\em any} quantum gate library, and for {\em any} cost metric.

\section{Conclusion}\label{sec:conc}

We have introduced quantum templates and demonstrated how they can 
be applied for quantum circuit simplification and level compaction.
Templates can be developed for any type quantum circuit, and
can be applied for various cost metrics (simple gate count,
weighted gate count, non-linear metrics). We implemented our
algorithms in C++ and demonstrated the
effectiveness of our approach using a variety of previously
published circuits. In our tests, we first target gate 
minimization and then compact the logic levels in the 
simplified circuit. In particular, we reduced the sizes and 
number of logic levels in the best known multiple control 
Toffoli gate quantum realizations (including multiple control 
Toffoli gates with negative controls) and in a number 
of arithmetic quantum circuits presented by previous authors.

\section*{Appendix}
Consistency of the template definition is based on the following 
four lemmas.

\vspace*{12pt}
\noindent
{\bf Lemma~1:} For any circuit
$G_0G_1...\;G_{m-1}$ realizing a quantum function $f$, circuit
$G^{-1}_{m-1}G^{-1}_{m-2}...\;G^{-1}_0$ is a realization for
$f^{-1}$.

\vspace*{12pt}
\noindent
{\bf Proof:}
This statement follows from the properties of matrix multiplication operation.
$\square$

\vspace*{12pt}
\noindent
{\bf Lemma~2:} For any rewriting rule
$G_1G_2...\;G_k \rightarrow G_{k+1}G_{k+2}...\;G_{k+s}$ its gates
satisfy the following:
$G_1G_2...\;G_kG^{-1}_{k+s}$ $G^{-1}_{k+s-1}...\;G^{-1}_{k+1} = I$,
where $I$ denotes the identity matrix (transformation).

\vspace*{12pt}
\noindent
{\bf Proof:}
The following set of equalities constructed using the rule
$GG^{-1}=I$ for a single gate $G$ proves the statement.
\begin{eqnarray*}
G_1G_2...\;G_k = G_{k+1}G_{k+2}...\;G_{k+s} \\
G_1G_2...\;G_kG^{-1}_{k+s}G^{-1}_{k+s-1}...\;G^{-1}_{k+1} = \\
= G_{k+1}G_{k+2}...\;G_{k+s}G^{-1}_{k+s}G^{-1}_{k+s-1}...\;G^{-1}_{k+1} \\
G_1G_2...\;G_kG^{-1}_{k+s}G^{-1}_{k+s-1}...\;G^{-1}_{k+1} = I.
\end{eqnarray*}
$\square$

\vspace*{12pt}
\noindent
{\bf Lemma~3:} For an identity
$G_0G_1...\;G_{m-1}$ and any parameter $p,\; 0 \leq p \leq m-1,$
$G_0G_1...G_{p-1} \rightarrow
G^{-1}_{m-1}G^{-1}_{m-2}...\;G^{-1}_p$ is a rewriting rule.

\vspace*{12pt}
\noindent
{\bf Proof:}
Proof of this statement follows from the previous one by renaming
the subscripts and listing the equalities in the reverse order.
$\square$

\vspace*{12pt}
\noindent
{\bf Lemma~4:} If $G_0G_1...\;G_{m-1} = I$, then
$G_1...\;G_{m-1}G_0 = I$.

\vspace*{12pt}
\noindent
{\bf Proof:}
The following proves the statement.
\begin{eqnarray*}
G_0G_1...\;G_{m-1} = I \\
G^{-1}_0G_0G_1...\;G_{m-1} = G^{-1}_0I \\
G_1...\;G_{m-1} = G^{-1}_0 \\
G_1...\;G_{m-1}G_0 = G^{-1}_0G_0 \\
G_1...\;G_{m-1}G_0 = I.
\end{eqnarray*} 
$\square$

\section*{Acknowledgements}

\noindent This work was supported by PDF and Discovery grants 
from the National Sciences and Engineering Research Council of Canada.

\end{document}